\documentclass[journal= aip,amsmath,amssymb, reprint]{revtex4-1}

\usepackage{hyperref}
\usepackage{booktabs}
\usepackage{tabularx}
\usepackage{comment}
\usepackage{graphics}

\usepackage{graphicx}

\usepackage[utf8]{inputenc}
\usepackage[T1]{fontenc}
\usepackage{mathptmx}
\usepackage{etoolbox}
\usepackage{float}
\makeatletter
\def\@email#1#2{%
 \endgroup
 \patchcmd{\titleblock@produce}
  {\frontmatter@RRAPformat}
  {\frontmatter@RRAPformat{\produce@RRAP{*#1\href{mailto:#2}{#2}}}\frontmatter@RRAPformat}
  {}{}
}%

\begin{document}

\author{Saleh Abdul Al} 
\affiliation{Lebanese International University - Beirut Campus Mouseitbeh - Mazraa, Beirut,  LB 146404, Lebanon}
\author{Abdul-Rahman Allouche} 
\affiliation{Institut Lumi\`ere Mati\`ere, CNRS, UMR  5306, Univ Lyon, Universit\'e Claude Bernard Lyon 1, F-69622 Villeurbanne, France}
\email{allouchear@univ-lyon1.fr} 

\begin{abstract}
Accurately predicting infrared (IR) spectra in computational chemistry using ab initio methods remains a challenge. Current approaches often rely on an empirical approach or on tedious anharmonic calculations, mainly adapted to semi-rigid molecules. This limitation motivates us to explore alternative methodologies. Previous studies explored machine-learning techniques for potential and dipolar surface generation, followed by IR spectra calculation using classical molecular dynamics. However, these methods are computationally expensive and require molecule-by-molecule processing. Our article introduces a new approach to improve IR spectra prediction accuracy within a significantly reduced computing time. We developed a machine learning (ML) model to directly predict IR spectra from three-dimensional (3D) molecular structures. The spectra predicted by our model significantly outperform those from density functional theory (DFT) calculations, even after scaling. In a test set of 200 molecules, our model achieves a Spectral Information Similarity Metric ($SIS$)  of 0.92, surpassing the value achieved by DFT scaled frequencies, which is 0.57. Additionally, our model considers anharmonic effects, offering a fast alternative to laborious anharmonic calculations.   Moreover, our model can be used to predict various types of spectra (Ultraviolet or Nuclear Magnetic Resonance for example) as a function of molecular structure. All it needs is a database of 3D structures and their associated spectra.
\end{abstract}

\title[]
  {Neural Network Approach for Predicting Infrared Spectra from 3D Molecular Structure}
  
\keywords{Machine Learning, Neural Network, Infrared Spectra}
\maketitle
\section{Introduction}
In computational chemistry, the accurate prediction of infrared (IR) spectra using ab initio methods remains a challenging endeavor. The inherent difficulty often necessitates either the application of empirical scaling to harmonic frequencies\cite{Trujillo22,Trujillo23} or engagement in a highly time-consuming anharmonic calculation, predominantly feasible with good accuracy only for semi-rigid molecules\cite{Bowman78,yagi00, Jung96,Carbonniere10,Nielsen51}. This limitation prompts the exploration of alternative methodologies to overcome the challenges associated with traditional approaches.

Several studies have sought to leverage Machine Learning techniques for generating potential and dipole surfaces, followed by the computation of IR spectra using classical molecular dynamic\cite{Gastegger17,Schutt21} or static\cite{Lam20,Laurens21} approaches. However, a prevalent issue lies in the fact that these spectra are typically calculated either at the harmonic approximation or anharmonic approximation levels, which proved to be effective only for semi-rigid molecular structures. This limitation underscores the need for a more versatile and accurate predictive model that extends its applicability to a broader range of molecular flexibilities.

Addressing this gap, a noteworthy avenue explored in previous research involved IR spectrum prediction using Simplified Molecular Input Line Entry System (SMILES) notation, using a 2D molecular graph\cite{McGill21} or Morgan-based fingerprints\cite{Kovacs20}. Nevertheless, the drawback in such approaches is the compromise on the fidelity of 3D molecular structures, raising questions about the reliability of predictions of their IR spectra.

{In light of these challenges, this paper proposes a novel approach to improve the accuracy of IR spectrum predictions. By combining the strengths of machine learning with a consideration of molecular structure, we aim to build a predictive model with superior accuracy compared to standard methods in the field.}  We note that in this work, our approaches will be applied to the infrared spectra of molecules. However, these approaches can easily be extended to other types of molecular spectra, such as Raman, NMR, or Optical-UV, for example.
 
\section{Computational Details and Methodology}
We employ message passing neural networks (MPNNs)\cite{Tang23,Unke19} to predict the infrared spectrum from molecular structure properties, such as atom symbols, 3D geometry, and atom masses. To train our model, a database of molecular structures and corresponding experimental infrared spectra is required. For structures, we utilized the Density Functional Theory (DFT)\cite{DFT99} method to optimize geometries. Our study incorporates, 4,896 experimental spectra in the gas phase, along with their DFT-optimized structures. After training, the model can predict the IR spectrum of any molecule based on its structure.

\subsection{Dataset}
Our goal is to reproduce experimental spectra from molecular structures ; therefore, we need an experimental database. Infrared spectra can be obtained experimentally in the gas phase or in condensed phases. In the condensed phase, the IR depends on the solvent system. For a given molecule, the spectrum in a pure component liquid film can differ from that in a pressed KBr pellet, in mineral oil suspension, or in a CCl4 solution. To obtain a good, accurate spectrum, it is necessary to train a model for each type of solvent. In this paper, we limited our study to spectra obtained in gas phase.  The experimental spectra were obtained from the National Institute of Standards and Technology (NIST)\cite{NISTDataBase}. This (commercial) dataset contains 5,228 infrared spectra of different compounds. The chemical structures are provided in the MOL-file format, using file names containing the CAS registry number of the compound. Using the CAS numbers, we retrieved 3D structures from the NIST WebBook site \cite{NISTWebBook}. We removed the compounds for which the 3D structures are not available. In a few cases, we encountered the same structure for different CAS numbers and removed them from the dataset. At the end, we obtained a dataset containing experimental infrared spectra and 3D structures for 4,896 molecules. Each spectrum is broadened by applying a Gaussian convolution, with a standard deviation of 10 $cm^{-1}$. Finally, Spectra processed to conform to a uniform data format at 2 $cm^{-1}$ intervals over the range 400-4000 $cm^{-1}$. The structures obtained from NIST Website, are optimized using B3LYP functional\cite{B3LYP88}, with def2-tzvp\cite{Weigend2005} as basis. The D3 dispersion with Becke-Johnson damping \cite{Grimme11} is added to  DFT energy. All DFT calculations are performed using Gaussian 16 software\cite{g16}. {Our dataset contains molecules with H, D, B, C, N, O, F, Si, P, S, Cl, Br, Sn, Te, I, and Hg atoms. The number of atoms per molecule ranges from 2 to 92. The histogram depicting this distribution is provided in the Supplementary Materials \ref{supmat}, Figure S1.}

\subsection{Neural Network model}
\begin{figure*}
    \centering
    \includegraphics[width=0.99\linewidth]{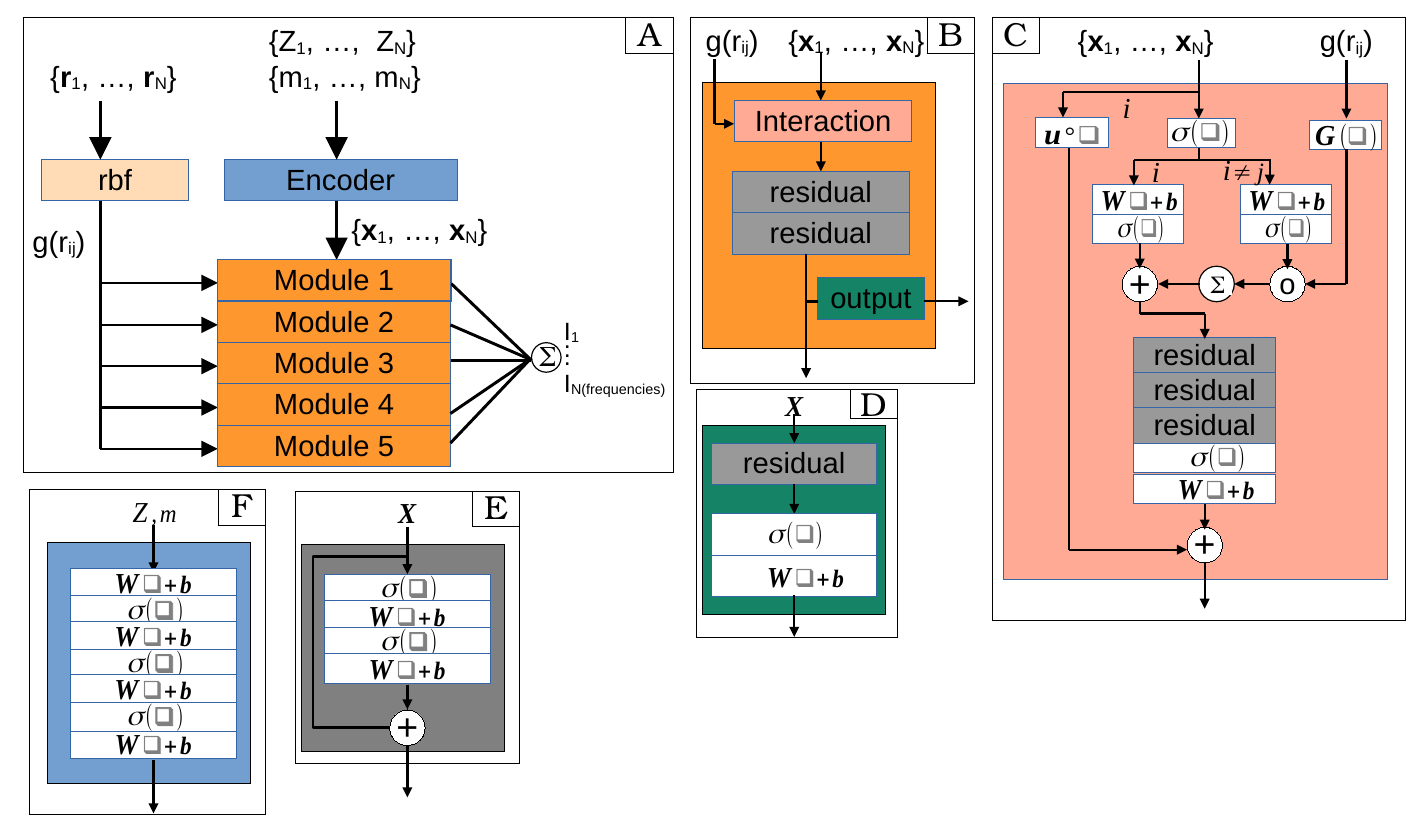}
    \caption{
    Overview over our Model architecture.\textbf{A}: The input nuclear charges $Z_i$ and nuclear masses $ m_i$ of $N$ atoms are transformed to feature vectors $ \mathbf {x_i} \in \mathbb {R} ^F (F=128) $ via an encoder block, and passed iteratively through a stack of five modules. From the Cartesian coordinates $ \mathbf {r_i} $, all pairwise distances are calculated and expanded in a set of K radial basis functions (rbf) forming the vectors  $   {g(r_{ij})}  \in \mathbb {R}^K (K=128) $ as additional inputs to each module. The output of all modules is summed to form the predictions of the intensities of the infrared spectrum. \textbf{B}: Each module transforms its input through an interaction block followed by two residual blocks. The computation then splits into two branches: One as input for next module, the second as  output block. \textbf{C}:  After passing through the activation function $\sigma $, the incoming features of the central atom $i$ and neighboring atoms $j$ split paths and are further refined through separate nonlinear dense layers.  The mask G ($g(r_{ij}))$ selects features of atoms $j$ based on their distance to atom $i$ and adds them to its features in order to compute the proto-message $\Tilde{v}$, which is refined through three residual blocks to the message $v$. After an additional activation and linear transformation, $v$ , which represents the interactions between atoms, is added to the gated feature representations $u\,^o x$. \textbf{D} :  An output block passes its input through one residual blocks  and a dense layer (with linear activation) to compute the final output of a module. \textbf{E}: Each residual block refines its input by adding a residual computed by a two-layer neural network. \textbf{F}: The encoder has 2 reals as input (Z and m for an atom) and a vector of F reals values as output. It is a sequence of four dense layers and three activation functions. { Finally, it is important to mention that we used the shifted softplus as the activation function, and we applied the softplus function to the output vectors.}
    }
    \label{fig:NNModel}
\end{figure*}

We implement an end-to-end learning architecture to predict the IR spectrum from a 3D molecular structure. Our model is similar to that of PhysNet one.\cite{Unke19} In PhysNet, the embedding layer is a vector for each $Z_i$ ($Z_i$ represents the atomic number of atom number i). {Since our dataset includes the D-isotope, the infrared spectrum might vary based on the atomic masses ($m_i$) of the molecule.} Consequently, we have replaced the embedding layer with an encoder (Figure \ref{fig:NNModel}.F) to represent the atomic number ($Z_i$) and atomic mass ($m_i$) of each atom by a vector. The input data of this encoder is a set of vectors of atomic numbers (Z) and also atomic masses in the molecule\cite{Herr19}. The complete architecture is schematically represented in Figure \ref{fig:NNModel}.
Our encoder is composed of 3 hidden layers, each comprising 128 nodes, with the output layer composed of a vector of $F=128$ values (F=the dimensionality of the feature space) values{, which is equal to the dimensionality considered in PhysNet}\cite{Unke19}.
This vector serves as features $x_i$ for the first module block.  This module (see Figure\ref{fig:NNModel}.B) couples the features $x_i$ of each atom i with the features $x_j$ of all atoms j within the cutoff distance $r_{\text{cut}}$ (=10 Bohr)\cite{Gastegger17} through an interaction block (Figure \ref{fig:NNModel}.C), using K=128 gaussian radial basis functions computed using the interatomic distances $r_{ij}$.The computed $x_i$ splits into two branches: one branch propagates the atomic features to the subsequent module, whereas the other branch forwards the features to an output block dedicated to compute the IR spectrum. We used 5 modules. Each output block takes $x_i$ as input and generates a vector of 1801 values as output, representing the intensities for the infrared spectrum for frequencies, uniformly ranging from 400 to 4000 cm$^{-1}$. 
{In an ab initio calculation, the vibration modes are initially computed, and subsequently, the spectrum is constructed through convolution to compare it with the measured spectrum. However, our approach involves directly predicting the spectrum without the need to calculate vibration modes. For each frequency (1801 values between 400 and 4000), we calculate the intensity, which directly provides the spectrum. This approach is crucial because our experimental database comprises spectra rather than collections of mode frequencies.}
\subsection{Training}
The parameters of the neural network are optimized by minimizing a loss function ($L$) using Adam\cite{Adam17} with a learning rate of $10^{-3}$ and a batch size of 32. We used the spectral information divergence (SID)\cite{Chang00} as the loss function $L$, defined by:

$$ SID(I_{p},I_{t}) = \sum_{i} I_{p,i} \ln \frac{I_{p,i}}{I_{t,i}} + I_{t,i} \ln \frac{I_{t,i}}{I_{p,i}} $$

where $I_{p}$ and $I_{t}$ are the predicted intensities and experimental ones, respectively. The learning rate is decayed by a factor of 0.9 if the validation loss plateaus.

The dataset is split into $N_{train}$, $N_{evaluation}$, and $N_{test}$ structures in the training, validation, and test sets respectively.
Starting from several random initial parameters, and random splitting of the dataset, we trained 12 models(see table \ref{table:performance}) with different $N_{train}$, $N_{evaluation}$, and $N_{test}$ values and different random seed. After training, the model that performed best on the validation set is selected. Since the validation set is used indirectly during the training procedure, the performance of the final models is always measured on a separate test set.
As a metric to compare two spectra, we used the Spectral Information Similarity Metric ($SIS$)\cite{McGill21}, given by :
$$ SIS(I_{p},I_{t}) = \frac{ 1 }{1+  SID(\tilde{I}_{p},\tilde{I}_{t})} $$ where $\tilde{I}$ is a Gaussian convolution of $I$  using a standard  deviation of $ 10\, cm^{-1}$. {The calculated $SIS$ is a single scalar value expressing the similarity between two spectra, where each spectrum is broadened by applying a Gaussian convolution, and each spectrum is normalized to sum all the spectrum absorbances to unity. $SIS$ provides an easily accessible measure of prediction quality and typically follows trends in prediction behavior.\cite{McGill21} }

As metrics, we can also use the standard statistic errors as  Root Mean Squared Deviation ($RMSD$ ) and Mean Absolute Error ($MAE$)  defined respectively by :
$$ 
RMSD(I_{p},I_{t}) =  
\frac{1}{N} \sqrt{\sum_{i=1}^{N} (\tilde{I}_{p}-\tilde{I}_{t})^2}$$
$$
MAE(I_{p},I_{t}) =  
\frac{1}{N} \sum_{i=1}^{N} |\tilde{I}_{p}-\tilde{I}_{t}|
$$
where $N$ is the number of intensities valuers.
{It's worth noting that $RMSD$ and $MAE$ can also be used as loss functions with similar performance to $SID$, though with a slight advantage for the latter.\cite{McGill21}.}

\section{Results and Discussions}
\subsection{Performance.} 
As observed in Table \ref{table:performance}, the average ($\bar{x}$) $SIS$ score for $SIS$ using the test set with the best parameters for the evaluation set is approximately 0.74. However, when employing an ensemble model composed of 12 individually trained submodels, the performance improves, maintaining an $SIS$ score of 0.81. This trend is consistent across median, standard deviation, $MAE$, and $RMSD$.
\begin{table}[H]
\centering
  \caption{Twelve models are fitted. For each model, the number of structures is given for the training, validation, and test parts. $\bar{x}$, $\tilde{x}$, and $\sigma$ represent the average, the median, and the standard deviation of $SIS$ values. MAE (mean absolute error) and RMSD (root-mean-square deviation) denote the mean absolute error and root-mean-square deviation of intensities, using the experimental values as a reference. The "Ensemble" model is constructed using the parameters of the 12 training models. The last line corresponds to scaled frequencies (0.975 for frequencies < 2000 cm$^{-1}$, 0.956 for frequencies > 3000 cm$^{-1}$, and  0.961 for other values) of harmonic DFT spectra.}
  \label{table:performance}
  \begin{tabular}{lllllllll}
    \hline
    Model       & $N_{train}$      & $N_{valid}$.      & $N_{test}$      & $\bar{x}$ & $\tilde{x}$      &  $\sigma$     & MAE        & RMSD \\
    \hline
    1	          &4096	&400	&400	&0.73	&0.77	&0.21	&0.10   &0.31 \\
    2	          &4096	&400	&400	&0.73	&0.79	&0.23	&0.11	&0.31 \\
    3	          &4096	&400	&400	&0.63	&0.66	&0.23	&0.15	&0.42 \\
    4	          &4069	&400	&400	&0.70	&0.73	&0.21	&0.12	&0.34 \\
    5	          &3584	&900	&412	&0.73	&0.78	&0.21	&0.11	&0.31 \\
    6	          &3584	&900	&412	&0.71	&0.77	&0.23	&0.12	&0.34 \\
    7	          &3584	&900	&412	&0.72	&0.77	&0.21	&0.11	&0.31 \\
    8	          &4496	&200	&200	&0.72	&0.78	&0.21	&0.10	&0.31 \\
    9	         &4496	&200	&200	&0.73	&0.81	&0.23	&0.10	&0.31 \\
    10	         &4496	&200	&200	&0.73	&0.79	&0.22	&0.10	&0.31 \\
    11	         &4496	&200	&200    &0.74	&0.79	&0.23	&0.10	&0.29 \\
    12	         &4496	&200	&200   &0.74	&0.81	&0.22	&0.11	&0.31 \\
    Ens.	&		&	   &      &0.81	&0.86	&0.16	    &0.08	&0.23 \\
    Ens./Ave	&		&	   &      &0.92	&0.95	&0.10	    &0.05	&0.14 \\
    DFT	&       &		&	   &0.57	&0.59	&0.20	        &0.16	&0.51 \\
    \hline
  \end{tabular}
\end{table}
The predicted spectra from any of the individual training models or the ensemble model surpass those predicted by scaled DFT spectra. During model training, we can also utilize the average parameters from all steps instead of selecting parameters that yield the best results in the evaluation set. Using these parameters from all 12 models, we construct another ensemble model (named Ens./Ave. in Table \ref{table:performance}). As indicated in the table, this model performs exceptionally well, with an $\bar{x}$ of 0.92 and an $\tilde{x}$ of 0.95.

\begin{figure}[H]
    \centering
    \flushleft
    \includegraphics[scale=0.5]{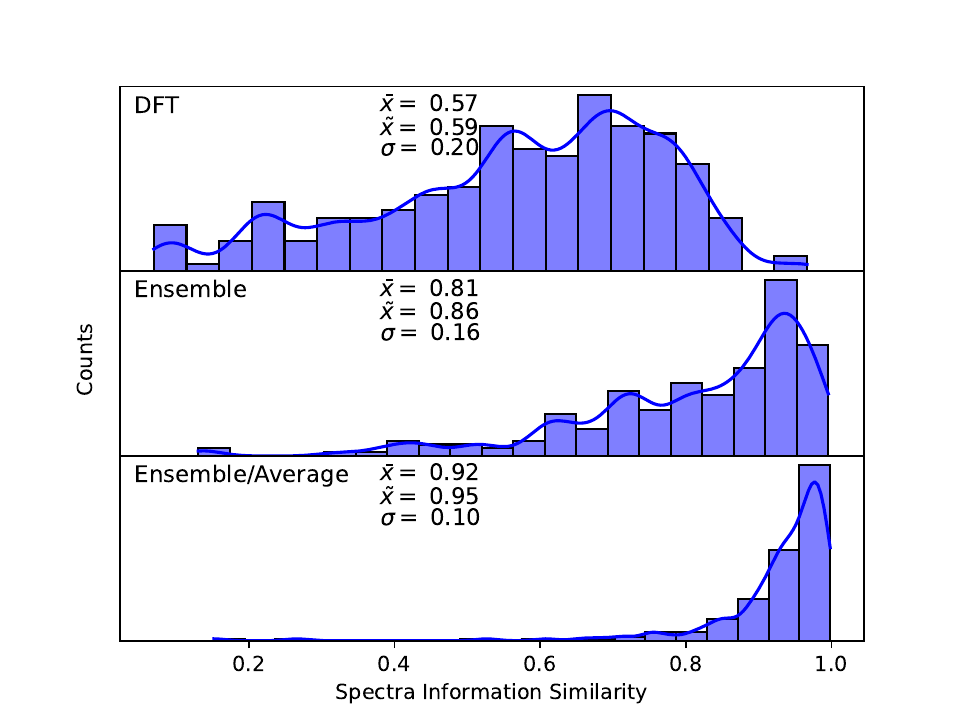}
    \caption{
   Distribution of the $SIS$ score (using experimental spectra as reference) for spectral predictions of the Ensemble models and DFT in a test of 200 structures. The average, the median values and standard deviation for
$SIS$ are provided as $\bar x$,$\tilde{x}$ and $\sigma$, respectively.
    }
    \label{fig:hist-ensemble}
\end{figure}

Figure \ref{fig:hist-ensemble} displays the distribution of $SIS$ scores for the predicted spectra using the two ensemble models and those calculated by DFT. It is evident that both models outperform DFT, with half of the spectra having an $SIS$ score above 0.59 for DFT, 0.86 for the ensemble model, and 0.95 for the Ensemble/Average model.

\begin{figure}[H]
    \centering
    \includegraphics[scale=0.5]{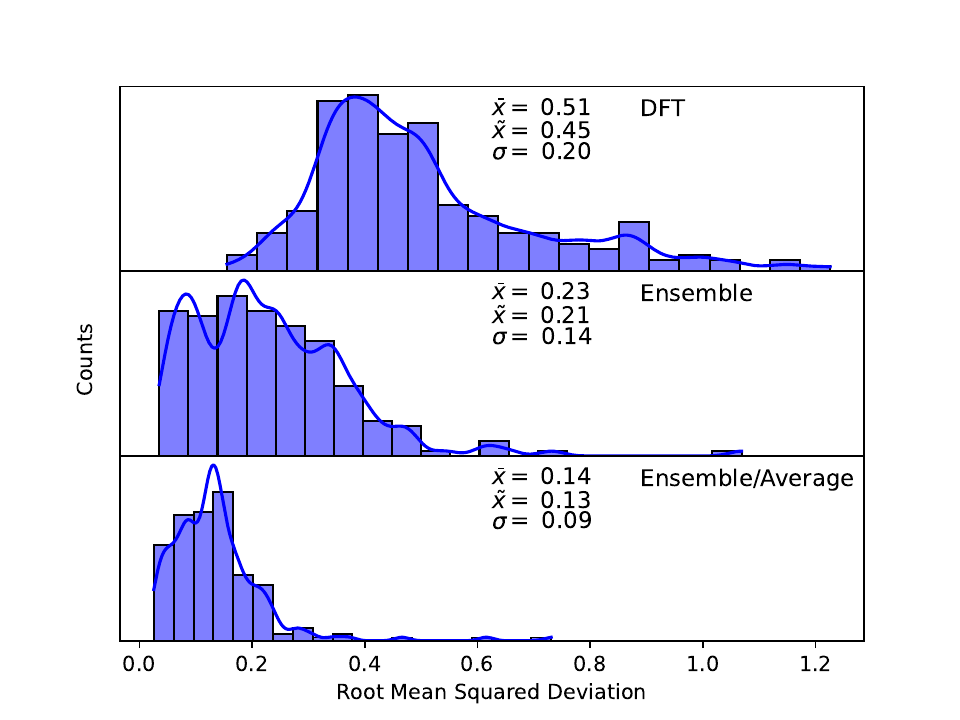}
    \caption{
   Distribution of the $RMSD$ (using experimental spectra as reference) for spectral predictions of the Ensemble models and DFT in a test of 200 structures. The average, the median values and standard deviation for
$RMSD$ are provided as $\bar x$,$\tilde{x}$ and $\sigma$, respectively.
    }
    \label{fig:hist-ensemble-rmsd}
\end{figure}

\begin{figure}[H]
    \centering
    \includegraphics[scale=0.5]{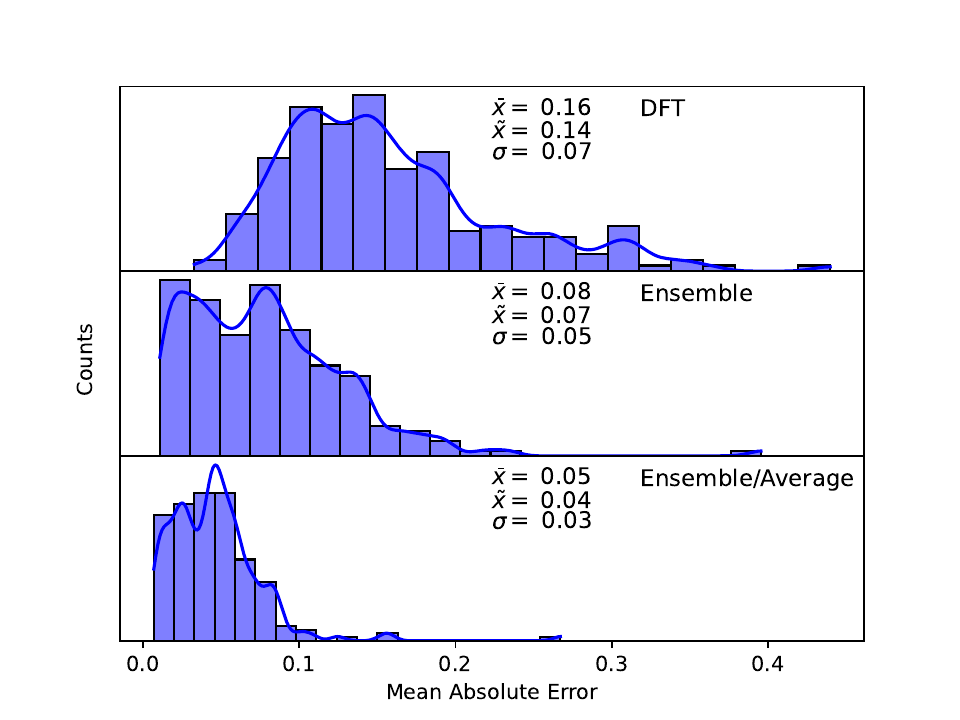}
    \caption{
   Distribution of the $MAE$ (using experimental spectra as reference) for spectral predictions of the Ensemble models and DFT in a test of 200 structures. The average, the median values and standard deviation for
$MAE$ are provided as $\bar x$,$\tilde{x}$ and $\sigma$, respectively.
    }
    \label{fig:hist-ensemble-mae}
\end{figure}

This trend can also be observed using $RMSD$ as a metric. In Figure \ref{fig:hist-ensemble-rmsd}, $RMSD$ varies from 0.2 to 1.2, with an average value of 0.51 for DFT. Meanwhile, it ranges from 0 to 0.6, with an average value of 0.2 for the Ensemble model, and is even smaller for the Ensemble/Average model, with an average value of only 0.14. The distribution of $MAE$ (Figure \ref{fig:hist-ensemble-mae}) closely resembles that of $RMSD$, with an average value of 0.05 and a very small standard deviation for the Ensemble/Average model. Thus, regardless of the metric used, it is evident that the two NN models (Ensemble and Ensemble/Average) outperform the DFT scale calculation.


\begin{figure*} 
\centering
    
    \includegraphics[width=0.85\textwidth]{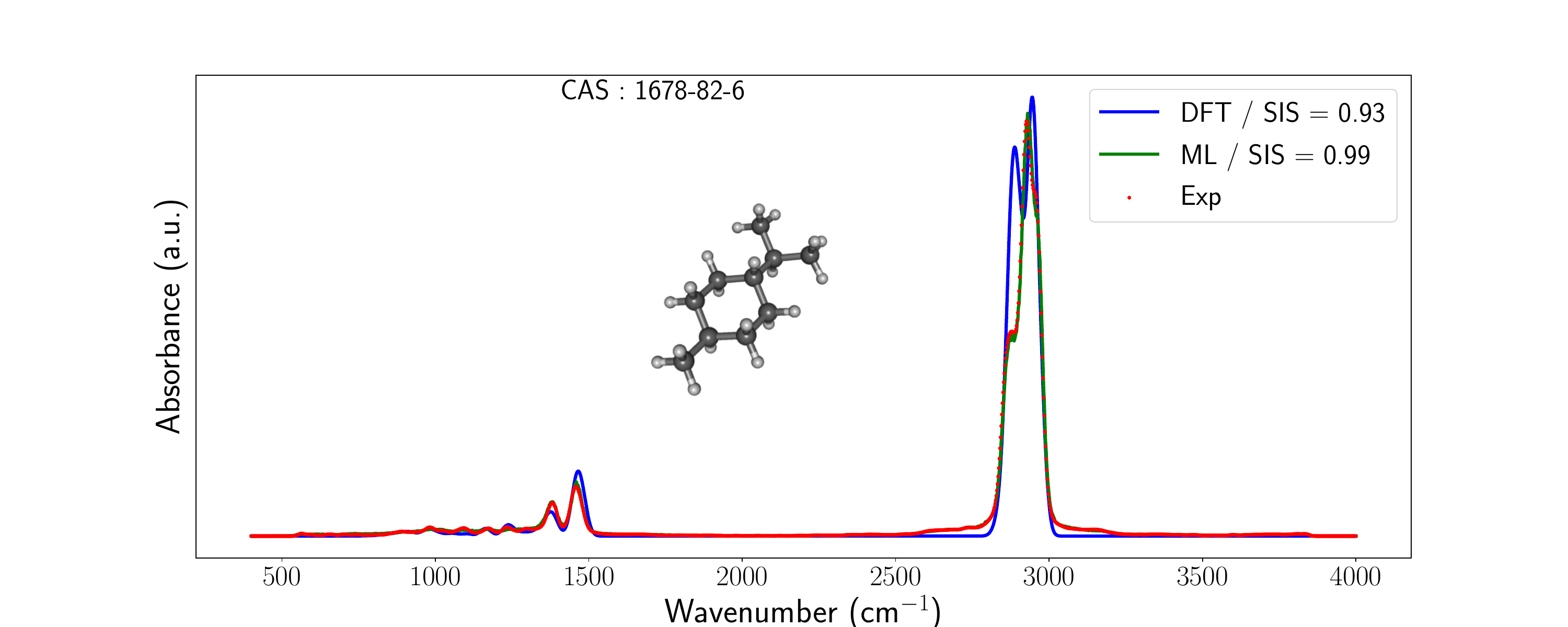}
  
 \caption{{Normalized spectra from experience, DFT and Ensemble/Average model for trans-1-Methyl-4-isopropylcyclohexane molecule.}}
 \label{fig:1678-82-6}
 \end{figure*}

\begin{figure*}  
   \centering
   \includegraphics[width=0.85\textwidth]{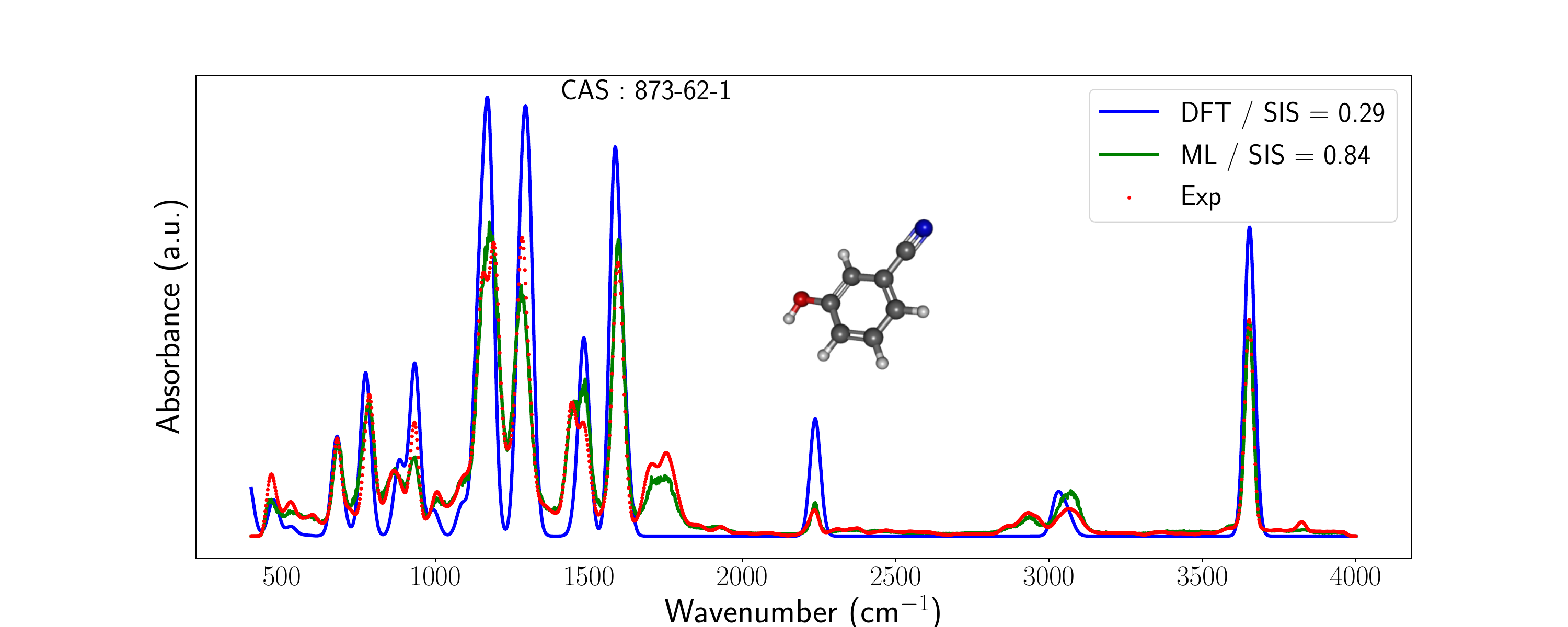}
 \caption{{Normalized spectra from experience, DFT and Ensemble/Average model for   Benzonitrile,3-hydroxy molecule.}}
  \label{fig:873-62-1}
 \end{figure*}

\begin{figure*}    
   \centering
   \includegraphics[width=0.85\textwidth]{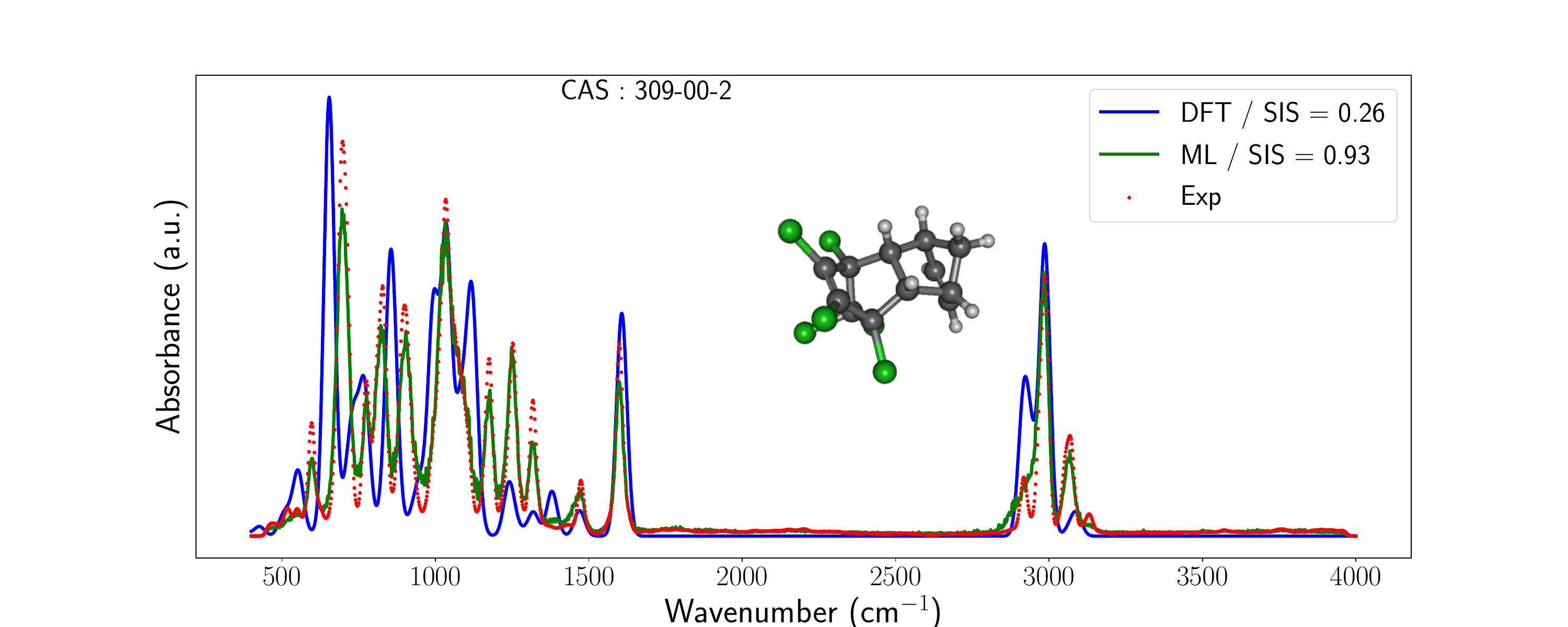}
    
 \caption{{Normalized spectra from experience, DFT and Ensemble/Average model for Aldrin molecule.}}
  \label{fig:309-00-2}
\end{figure*}

Examples of spectra at different percentile locations (for DFT) in the performance distribution are provided in Figures \ref{fig:1678-82-6},  \ref{fig:873-62-1}, \ref{fig:309-00-2}. These examples illustrate that the quality of predicted DFT spectra decreases with the complexity (types of atoms) of the molecule, while the quality of spectra predicted by the machine learning model remains consistent.
 
We observe that, in the DFT spectra, several peaks are not predicted correctly, while our models accurately predict both their positions and intensities, for both major and most minor peaks.
The pronounced distinction is particularly evident in high frequencies, where the influence of the anharmonic effect for stretching-type modes is very important. Consequently, in scaled DFT calculation, a smaller scaling factor is employed for high frequencies in comparison to low frequencies. Despite the uniformity of the correction factor across all stretching modes, it is not possible to establish a direct correlation between anharmonicity and bond type (e.g., C-H, O-H, NH, C-Cl). Consequently, it is compelling to analyze the performance of our model separately for the two frequency ranges: high and low.

\begin{figure}[h!]
    \centering
    \includegraphics[scale=0.5]{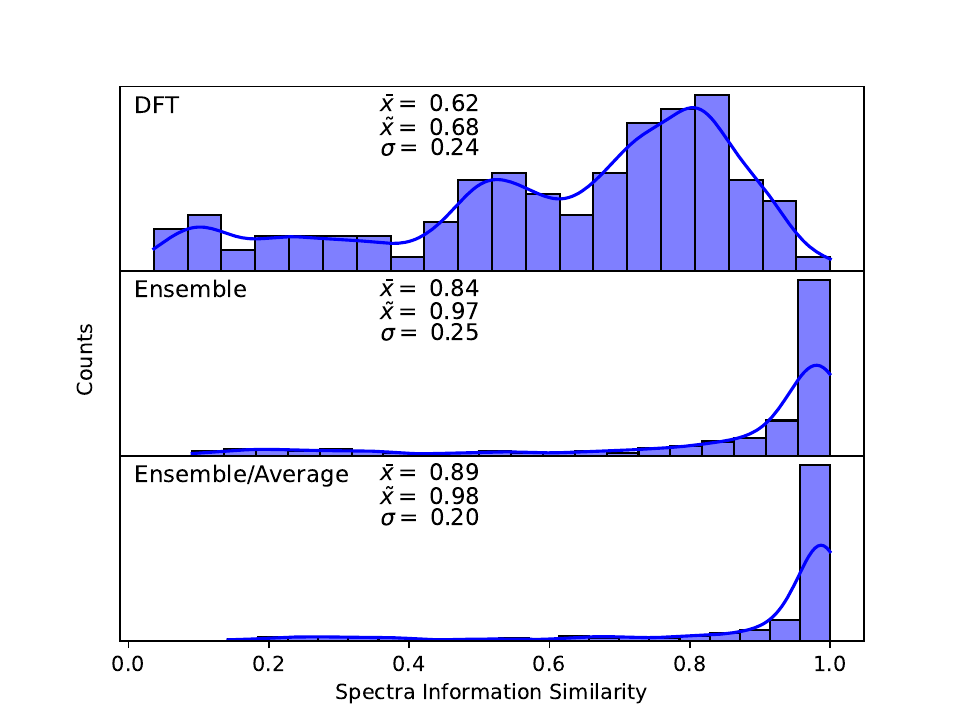}
    \caption{
Distribution of the $SIS$ scores (using experimental spectra as reference) for spectral predictions of the Ensemble models and DFT in a test of 200 structures, using frequencies greater than 2000 cm$^{-1}$. The average, median values, and standard deviation for $SIS$ are provided as $\bar{x}$, $\tilde{x}$, and $\sigma$, respectively.
    }
    \label{fig:hist-ensemble_sup2000}
\end{figure}

\begin{figure}[H]
    \centering
    \includegraphics[scale=0.5]{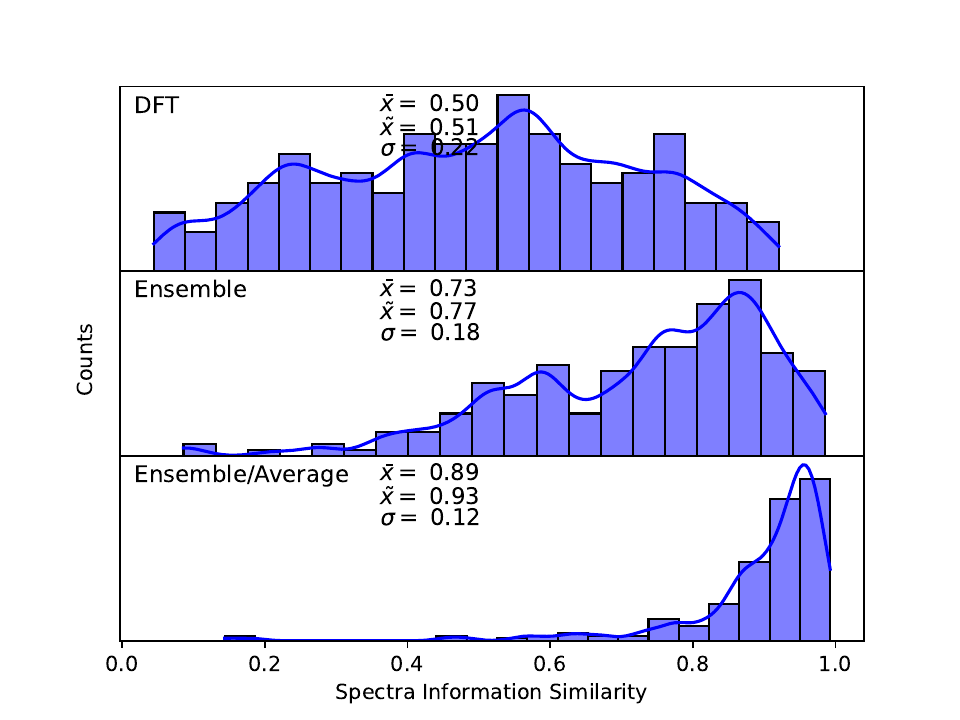}
    \caption{
   Distribution of the $SIS$ score (using experimental spectra as reference) for spectral predictions of the Ensemble models and DFT in a test of 200 structures, using frequencies lower 2000 cm$^{-1}$. The average, the median values and standard deviation for
$SIS$ are provided as $\bar x$,$\tilde{x}$ and $\sigma$, respectively.
    }
    \label{fig:hist-ensemble_inf2000}
\end{figure}

In Figures \ref{fig:hist-ensemble_sup2000} and \ref{fig:hist-ensemble_inf2000}, we present the distribution of the $SIS$ score for high and low frequencies. It is observed that the performance of DFT in the low-frequency range closely resembles that of the Ensemble model, while the reliability of the Ensemble/Average model is notably higher. However, for high-frequencies, both the Ensemble and Ensemble/Average models demonstrate considerably greater precision compared to scaled DFT calculations. This observation underscores the capacity of our model to incorporate nuances associated with different types of bonds (C-H, O-H, N-H, C-Cl etc.) and anharmonicity in interpreting IR spectra.

{
The number of atoms per molecule in our dataset ranges from 2 to 92. It is interesting to study the performance across different molecule sizes. The distributions of $SIS$ along the size of molecules are given in the Supplementary Materials (see \ref{supmat}, Figure S1). The mean number of atoms is 23. In our test dataset, there are 96 molecules with fewer than 23 atoms per molecule and 104 molecules with more than 23 atoms. The average $SIS$ values are 0.51, 0.77, and 0.91 for the first part, and 0.62, 0.85, and 0.94 for the second part, for DFT scaled frequencies, Ensemble, and Ensemble/Average models, respectively (see \ref{supmat}, Figures S2\&S3). It is clear that our model outperforms the DFT scaled frequencies calculation for both small and large molecules.
}

{
\subsection{Comparison with a neural network model based on SMILES}
McGill et al. \cite{McGill21} developed Chemprop-IR, a model based on 2D structure, using SMILES of the molecule as input for a neural network model. To compare the performance of our model to Chemprop-IR, we built the SMILES from CAS numbers using the CIRpy code \cite{CIRpy}, a Python interface for the Chemical Identifier Resolver (CIR) by the CADD Group at the NCI/NIH. We used the experimental averaged model fitted of Chemprop-IR \cite{ChempropIR}. The distribution of the $SIS$ score (using experimental spectra as reference) for spectral predictions of our Ensemble models, DFT, and Chemprop-IR in a test set of 200 structures is given in the Supplementary Materials \ref{supmat}, Figure S4. We noticed that the performance of Chemprop-IR is similar to DFT scaled frequencies, with an average $SIS$ of 0.57, and is outperformed by our models, which achieved 0.81 and 0.92 with Ensemble and Ensemble/Average, respectively.
}

{
\subsection{Hyperparameters}
The architecture of our model, depicted in Figure \ref{fig:NNModel}, can be adjusted through hyperparameters: F, K, the number of modules, and the number of residuals within each block.  Increasing the width and depth of the neural network generally leads to enhanced performance. However, since an increased width and depth is also associated with a higher computational cost, a compromise needs to be
found. While it would be possible to optimize hyperparameters, for example via a grid search, it
was found that this is not necessary for good performance
across different tasks. For simplicity, all models used in this work share the same architecture using the same hyperparameter of physNet model \cite{Unke19}. For the encoder block, we used 3 hidden layers. Moreover, we experimented with training models where K=F=64 and K=F=128, using all molecules in our dataset and the same number of epochs for training. We found (see Supplement materials \ref{supmat}, Figure S8) that performance with K=F=64 was already very good, with the best performance observed for K=F=128. Thus, we can conclude that F=K=128 represents a good compromise between performance and computational time.
We used a cutoff of 10 Bohr ($\approx$ 5.3 Å) for all previous results. In the literature, the cutoff ranges from 4.0 Å \cite{Gastegger17} to 10 Å \cite{Unke19}, frequently falling between 8 ($\approx$ 4.2 Å) and 10 Bohr ($\approx$ 5.3 Å) \cite{Ko2021}. We note here that in our message passing model, we used 5 modules, so the interaction is not limited to neighboring atoms defined by the cutoff. The range of interaction is controlled not only by the cutoff, but also by the number of passing steps (see \cite{Behler2021} for more details).
}

\subsection{Computational Time}
The computational cost for machine learning models is a predominant factor during model training, while prediction tasks are relatively fast and inexpensive by comparison. In scaled DFT calculations, the computational cost is mainly due to the calculation of harmonic frequencies. For example, the computation of harmonic DFT for the 4,896 molecules of our database requires 29,086 hours of CPU time (on an AMD EPYC 7302), whereas predictions for all these molecules take only 1.6 hours using our neural network model on the same machine and only 0.6 hour on GPU (NVIDIA V100).  Consequently, our model demonstrates both superior speed and precision. Note that explicit calculation methods for anharmonicity are significantly more expensive than harmonic calculations. Therefore, our NN model is much faster than these methods.

{
\subsection{Potential Enhancements}
The augmentation of the number of molecules in the database is anticipated to positively influence the model’s performance. Currently, we have utilized spectra from 4896 molecules. It is highly likely that there exist molecules in nature with interatomic interactions not included in our database. Consequently, our model may not accurately predict the spectra of these molecules. Expanding our database would undoubtedly be advantageous. However, such spectra must be experimental gas-phase spectra, which limits the feasibility of enlarging the database size.
}

\section{Conclusion}
The utilization of machine learning (ML) models has proven to be highly effective in predicting infrared (IR) spectra from molecular structures with remarkable accuracy. The precision achieved by ML surpasses that of scaled Density Functional Theory (DFT) calculations. Moreover, the ML model demonstrates the capability to incorporate anharmonic effects into the spectrum prediction based on the molecular structure. This innovation not only enhances predictive accuracy but also circumvents the time-consuming nature of explicit anharmonic calculations, such as GVPT2\cite{Martin95} and VCI\cite{Carbonniere10} methods. Overall, the application of ML in this context not only refines spectral predictions but also offers a more efficient alternative to traditional computational methods.
We note that our model architecture is inherently versatile and capable to take into account various interaction effects on spectra. In conclusion, our model exhibits the potential to be trained for the prediction of diverse spectra types, including Raman , NMR and Optical-UV based on molecular structure. This necessitates the construction of a comprehensive dataset comprising molecular structures and their corresponding spectra.

\section*{Supplementary Materials}{\label{supmat}The following supporting information can be downloaded at: \\
\begin{tabular}{ll}
\textbf{Name}   & \textbf{Description} \\
cas-molnames.txt   & CAS numbers and formula for molecules \\
SFigures  & Supllementary materials figures \\
\end{tabular}

}

\section* {Data availability}{
The NNMol-IR\cite{NNMol-IR} software developed in this study, will be made available (upon acceptance of the paper) through a public repository on GitHub.
{
The data that support the findings and fitted models will be deposed in \href{https://zenodo.org/}{Zenodo}.
}
However, it's important to note that the experimental data \cite{NISTDataBase} used in training these models are restricted by copyright held by the respective organizations and cannot be shared directly
}

\section*{Authors contributions}{
\textbf{Saleh Abdul Al}: Conceptualization (equal);  Formal analysis (equal); Writing – original draft (equal); Writing – review \& editing (equal). 
\textbf{Abdul-Rahman Allouche }: Software (lead);  Formal analysis (equal); Writing – original draft (equal); Writing – review \& editing (equal).
}

\acknowledgments{
This work was made possible only through the generous allocation of computer time provided by the 'Centre de calcul CC-IN2P3' in Villeurbanne, France.
}

\section*{conflicts of interest}
{The authors have no conflicts to disclose.}
\vspace{6pt} 


\section*{Abbreviations}
The following abbreviations are used in this manuscript:\\
\noindent 
\begin{tabular}{@{}ll}
DFT & Density functional theory \\
$MAE$ & Mean Absolute Error\\
$MPNN$ & Message passing neural networks\\
$RMSD$ &  Root Mean Squared Deviation  \\
SMILES & Molecular Input Line Entry System \\
$SID$ &  Spectral information divergence \\
$SIS$ & Spectral Information Similarity Metric\\
\end{tabular}
\\


\bibliography{mybib.bib}
\end{document}